\documentclass[conference]{IEEEtran}
\usepackage[dvipsnames]{xcolor}

\usepackage{graphicx}

\usepackage{tikz}
\usepackage[most]{tcolorbox}
\usepackage{gensymb}
\usepackage{todonotes}
\usepackage{amsmath,amsfonts}
\usepackage{algorithm}
\usepackage[noend]{algpseudocode}
\usepackage[symbol]{footmisc}

\usepackage{tikz}
\usepackage{amsmath}
\usepackage{multirow}
\usepackage{booktabs}

\newcommand*\circled[1]{\tikz[baseline=(char.base)]{
            \node[shape=circle,fill,inner sep=1pt] (char) {\textcolor{white}{#1}};}}
\usepackage{xcolor}
\usepackage{pgfplots}
\usepackage{tikz}
\pgfplotsset{compat=1.8}
\pgfplotsset{
    width=\textwidth,
}
\usepackage{lipsum}
\usepackage{pgfplotstable}
\usetikzlibrary{pgfplots.groupplots}

\usepackage{tikz}

\definecolor{codegreen}{rgb}{0,0.6,0}
\definecolor{codegray}{rgb}{0.5,0.5,0.5}
\definecolor{codepurple}{rgb}{0.58,0,0.82}
\definecolor{mGreen}{rgb}{0,0.6,0}
\definecolor{mGray}{rgb}{0.5,0.5,0.5}
\definecolor{mPurple}{rgb}{0.58,0,0.82}
\definecolor{backcolour}{rgb}{0.95,0.95,0.92}

\definecolor{RYB1}{RGB}{80, 99, 42}
\definecolor{RYB2}{RGB}{215, 227, 191}
\definecolor{RYB3}{RGB}{198, 187, 174}
\definecolor{RYB4}{RGB}{146, 205, 220}
\definecolor{RYB5}{RGB}{238, 144, 34}
\definecolor{RYB6}{RGB}{142, 172, 59}

\pgfplotscreateplotcyclelist{colorbrewer-RYB}{
{RYB1!50!black,fill=RYB1},
{RYB2!50!black,fill=RYB2},
{RYB3!50!black,fill=RYB3},
{RYB4!50!black,fill=RYB4},
{RYB5!50!black,fill=RYB5},
{RYB6!50!black,fill=RYB6},
}

\usepackage{lscape} 
\usepackage{pdflscape}
\usepackage{afterpage}
\usepackage{capt-of}
\usepackage{listings}
\usepackage{float}
\usepackage{array,multirow,graphicx}
\usepackage{caption}
\usepackage{subcaption}
\usepackage{soul}
\usepackage{pifont}

\definecolor{ggreen}{HTML}{2CC225}
\definecolor{yyellow}{HTML}{C2C80A}
\definecolor{bbrown}{HTML}{8e4603}

\usepackage{lscape} 
\usepackage{pdflscape}
\usepackage{afterpage}
\usepackage{capt-of}
\usepackage{listings}
\usepackage{float}
\usepackage{array,multirow,graphicx}
\usepackage{caption}
\usepackage{subcaption}
\usepackage{soul}
\usepackage{graphicx}
\usepackage{pifont} 
\usepackage{pifont}
\usepackage{amssymb} 
\definecolor{deepgreen}{RGB}{0,100,0}

\usepackage{xcolor}

\definecolor{codegreen}{rgb}{0,0.6,0}
\definecolor{codegray}{rgb}{0.5,0.5,0.5}
\definecolor{codepurple}{rgb}{0.58,0,0.82}
\definecolor{mGreen}{rgb}{0,0.6,0}
\definecolor{mGray}{rgb}{0.5,0.5,0.5}
\definecolor{mPurple}{rgb}{0.58,0,0.82}
\definecolor{backcolour}{rgb}{0.95,0.95,0.92}

\lstdefinestyle{CStyle}{
    commentstyle=\color{mGreen},
    keywordstyle=\color{magenta},
    numberstyle=\tiny\color{mGray},
    stringstyle=\color{mPurple},
    basicstyle=\sffamily\footnotesize,
    frame=lrtb,
    breakatwhitespace=false,         
    breaklines=true,                 
    captionpos=b,                    
    keepspaces=true,                 
    numbers=left,                    
    numbersep=5pt,                  
    showspaces=false,                
    showstringspaces=false,
    showtabs=false,                  
    tabsize=2,
    language=C
}

\lstdefinestyle{CStyle1}{
    commentstyle=\color{mGreen},
    keywordstyle=\color{magenta},
    numberstyle=\tiny\color{mGray},
    stringstyle=\color{mPurple},
    basicstyle=\sffamily\footnotesize,    frame=lrtb,
    breakatwhitespace=false,         
    breaklines=true,                 
    captionpos=b,                    
    keepspaces=true,                 
    numbers=left,                    
    numbersep=5pt,                  
    showspaces=false,                
    showstringspaces=false,
    showtabs=false,                  
    tabsize=2,
    language=C
}

\lstdefinestyle{mystyle}{
    commentstyle=\color{codegreen},
    keywordstyle=\color{magenta},
    numberstyle=\tiny\color{codegray},
    stringstyle=\color{codepurple},
    basicstyle=\sffamily\footnotesize,
    breakatwhitespace=false,         
    breaklines=true,                 
    captionpos=b,                    
    keepspaces=true,                 
    numbers=left,                    
    numbersep=5pt,                  
    showspaces=false,                
    showstringspaces=false,
    showtabs=false,                  
    tabsize=2,
    language=C
}
\lstdefinestyle{trans}{
    commentstyle=\color{codegray},
    numberstyle=\tiny\color{codegray},
    stringstyle=\color{codepurple},
     basicstyle=\sffamily\footnotesize,
    frame=lrtb,
    breakatwhitespace=false,         
    breaklines=true,                 
    captionpos=b,                    
    keepspaces=true,                 
    numbers=left,                    
    numbersep=5pt,                  
    showspaces=false,                
    showstringspaces=false,
    showtabs=false,                  
    tabsize=2,
     language=[x86masm]Assembler,  escapeinside={\%*}{*)},   
     }     

\ifCLASSINFOpdf
\else
\fi

\begin{document}
%


\title{ChipletQuake: On-die Digital Impedance Sensing for Chiplet and Interposer Verification}


\author{\IEEEauthorblockN{Saleh Khalaj Monfared}
\IEEEauthorblockA{
Worcester Polytechnic Institute\\Worcester,MA\\
skmonfared@wpi.edu}
\and
\IEEEauthorblockN{Maryam Saadat Safa}
\IEEEauthorblockA{
Worcester Polytechnic Institute\\Worcester,MA\\
msafa@wpi.edu}
\and
\IEEEauthorblockN{Shahin Tajik}
\IEEEauthorblockA{Worcester Polytechnic Institute\\Worcester,MA\\
stajik@wpi.edu}}


%


\maketitle

\begin{abstract}

The increasing complexity and cost of manufacturing monolithic chips have driven the semiconductor industry toward chiplet-based designs, where smaller and modular chiplets are integrated onto a single interposer. While chiplet architectures offer significant advantages, such as improved yields, design flexibility, and cost efficiency, they introduce new security challenges in the horizontal hardware manufacturing supply chain. These challenges include risks of hardware Trojans, cross-die side-channel and fault injection attacks, probing of chiplet interfaces, and intellectual property theft.
To address these concerns, this paper presents \textit{ChipletQuake}, a novel on-chiplet framework for verifying the physical security and integrity of adjacent chiplets during the post-silicon stage. By sensing the impedance of the power delivery network (PDN) of the system, \textit{ChipletQuake} detects tamper events in the interposer and neighboring chiplets without requiring any direct signal interface or additional hardware components. Fully compatible with the digital resources of FPGA-based chiplets, this framework demonstrates the ability to identify the insertion of passive and subtle malicious circuits, providing an effective solution to enhance the security of chiplet-based systems. To validate our claims, we showcase how our framework detects Hardware Trojan and interposer tampering.


\end{abstract}


%
\IEEEpeerreviewmaketitle

 \section{Introduction}
The increasing complexity of monolithic chips has led to skyrocketing costs and significant technical challenges over the last decade. 
As semiconductor manufacturers push towards smaller nodes, the sophistication of scaling down components while maintaining performance, power efficiency, and cost becomes more difficult. 
On the other hand, manufacturing large-scale System-On-Chips (SoC) brings significant challenges regarding post-silicon testing procedures. 
In response, the industry has shifted towards multi-chip module (MCM) designs, where multiple smaller chiplets are integrated onto a single interposer. 
Such heterogeneous packaging approaches offer several advantages, such as allowing the combination of different process technologies, improving yields by using smaller, modular dies, and enhancing scalability. 

Contrary to monolithic ICs, creating systems from separately produced components creates security issues, e.g., the possibility of die-swapping, susceptibility to interposer probing, or tampering.
In a zero-trust security model, a chiplet should be able to verify other chiplets. 
Verification schemes such as delay-based PUFs between chiplets have been developed, where the signal propagation delays through the interposers are considered as fingerprints~\cite{deric2022know}.
However, there might be no direct signal connection between the verifier and prover chiplets to realize such PUFs.
Moreover, there could be sophisticated probing or tampering attacks on chiplets interfaces~\cite{deric2024evaluating}, which enable direct eavesdropping from the interposer wires.

\begin{figure}[t]
     	\centering \noindent
     	\includegraphics[width=0.5\textwidth]{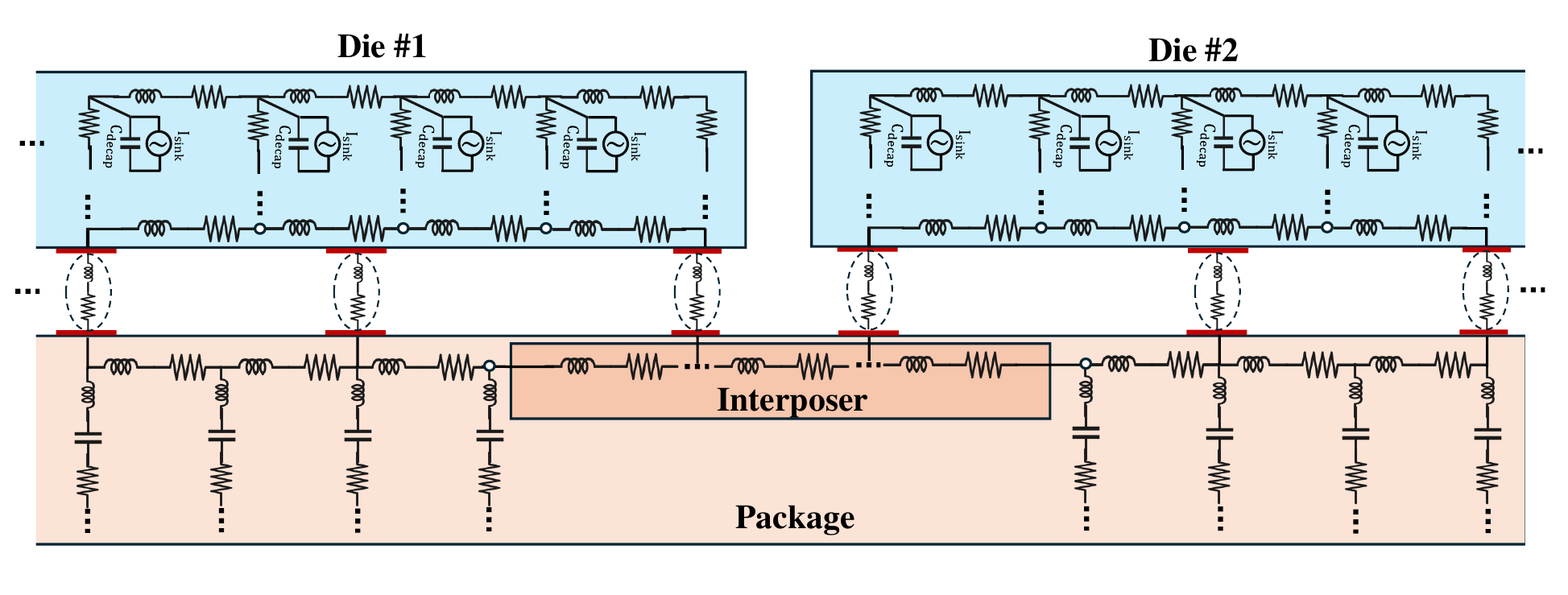}
     	\caption{A simple circuit model of a PDN and interposer inspired by~\cite{hossen2022analysis}.}
     	\label{pdn}
        \end{figure}

In MCMs, multiple chiplets and interposers are interconnected through a shared power distribution network (PDN).  Fig.~\ref{pdn} shows a simple model of a PDN and interposer.
The PDN distributes power from a central source to each chiplet. 
There have been a few attempts in the literature~\cite{zhang2023sipguard} to include self-contained sensors on one of the chiplets to detect anomalies in adjacent chiplets.
On digital ICs and field-programmable gate arrays (FPGAs), these sensors take the form of delay-based circuits such as on-chip ring-oscillators and time-to-digital converters.
If all goes well, any anomalies in running applications will then affect the sensitive timing behavior of these sensors.
However, such passive sensing methods have led to low-precision measurements and noisy behavior. Therefore, advanced machine learning methods are needed to obtain acceptable classification accuracy.
Moreover, all these solutions only demonstrate the detection of specific anomaly behavior in running software and are not applicable to physical tamper events, such as dormant hardware Trojans. 
Hence, it remains open if we can have a sensor on a trusted chiplet that can physically verify its environment beyond itself, from the interposer to neighboring chiplets, in a unified manner. 


\begin{figure*}[t!]
           \centering
        \begin{subfigure}{.69\textwidth}
         \includegraphics [width=\textwidth]{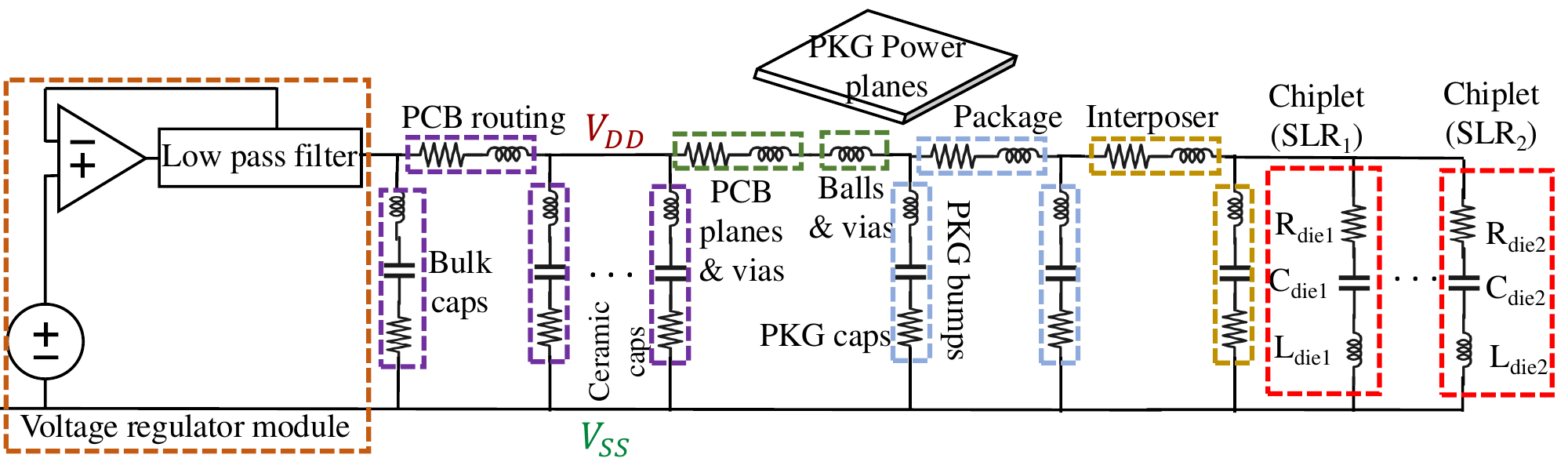}
                \caption{}
               \label{subfig:RLC_chip}
        \end{subfigure}
    \hspace{0.5pt}
        \begin{subfigure}{.29\textwidth}
         \includegraphics [width=\textwidth]{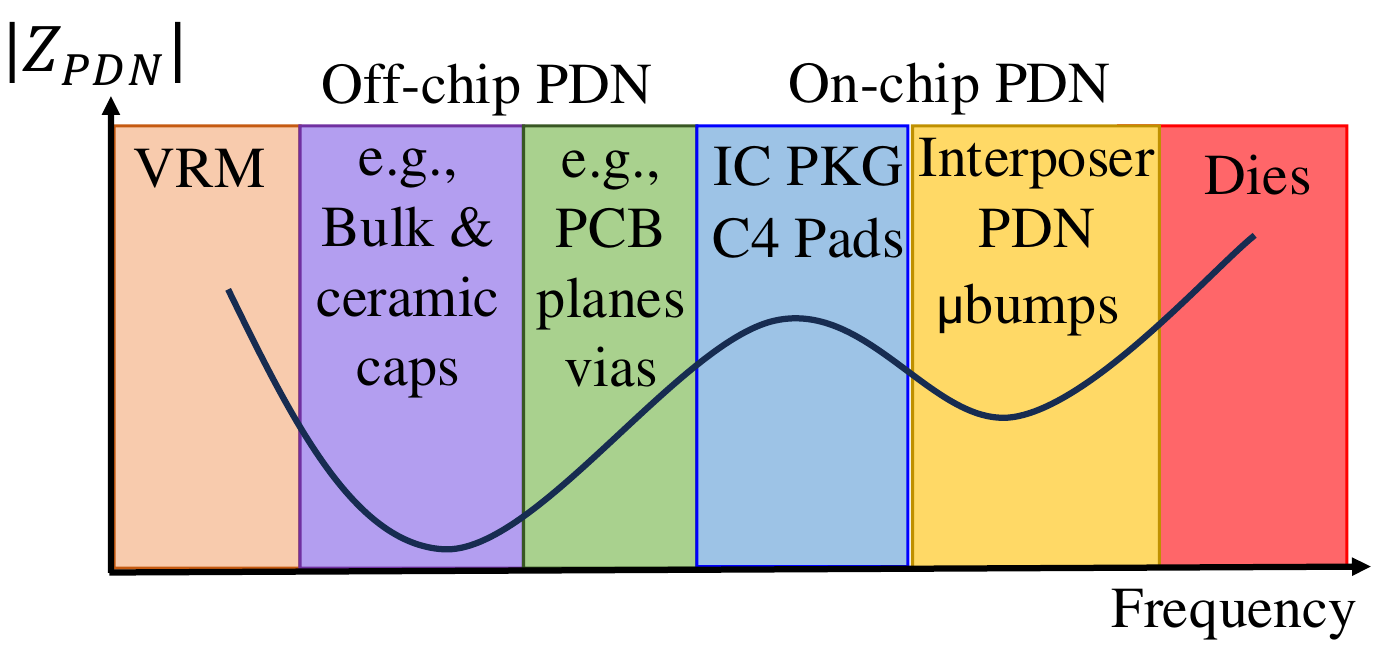}
                \caption{}
               \label{subfig:freq_disp}
               \vspace{7pt}
                 \end{subfigure}
	\caption{(a) Equivalent RLC circuit model of the power distribution network of the PCB and chip. (b) Contribution of different parts of the PDN to the impedance over frequency~\cite{mosavirik2023silicon}.}
   	\label{fig:RLC_Model_freq_disp}
 \end{figure*}

\textbf{Contribution.} In this work, we show that circuit-level elements (in terms of their hardware placements and routing) can be potentially fingerprinted in a PDN-shared heterogeneous system that is organized with multiple chiplets in a single package.
Our method relies on the fact that physical modifications (regardless of their physical size, activation, or action characteristics) alter the impedance of the shared PDN.
Therefore, characterizing the impedance can lead to the detection of the tamper events.
Although such PDN-based signatures are studied for chiplet fingerprinting as a possible attack vector in multi-tenant cloud FPGAs~\cite{zhu2023pdnsig}, we aim to show that chiplet and interposer monitoring is possible for security verification.
For this purpose, we present a systematic frequency-sweeping mechanism that can identify the integrity of the internal circuitry of the chiplets and the corresponding interposer interconnection circuits.
\noindent In summary, the key contributions of this manuscript are as follows:
\begin{itemize}
\item This article is the first work to introduce a fully-digital and spectral-assisted method to accurately ensure the hardware integrity of chiplets. 

\item We design a framework, namely~\textit{ChipletQuake}, which provides efficient and effective hardware verification of adjacent chiplets and host interposer. 

\item We implement~\textit{ChipletQuake} on a high-end chiplet-based FPGA system and perform extensive experiments against real-world hardware Trojans, and showcase the applicability of our framework. 

\end{itemize}

\section{Technical Background}
\subsection{Power Delivery Network (PDN)}
The Power Delivery Network (PDN) is critical in providing a stable and low-noise voltage supply to the electronic components on a PCB, spanning from the voltage regulator module (VRM) to the power rails on the chip.
The PDN is made up of both off-chip and on-chip elements, such as bulk capacitors, PCB routing, ceramic capacitors, PCB planes, vias, package bumps, on-chip power planes, and transistor capacitance. In the case of multi-die systems, interposer, and its corresponding $\mu$bumps, inter-die interconnections are also part of the PDN.
Each of these components contributes to the PDN’s impedance across different frequency regimes.

At low frequencies, the impedance of the PDN is primarily governed by the voltage regulator and off-chip components. 
As the frequency increases, the impedance behavior changes significantly, with the on-chip components contributing more to the PDN impedance at higher frequencies. This shift is largely due to parasitic inductance inherent in each capacitor, which affects their behavior at different frequencies.
At higher frequencies, capacitors exhibit a resonance phenomenon caused by the parasitic inductance in their metals.
Beyond this resonance frequency, the capacitors behave like open circuits, drastically reducing their impact on the PDN’s impedance.
Smaller capacitors, with lower parasitic inductance, resonate at higher frequencies, meaning their influence on the PDN impedance diminishes as the frequency increases.
Consequently, at very high frequencies, the impedance is dominated by interposer and on-chip structures, which are characterized by smaller dimensions.

The on-chip PDN behavior is modeled using an equivalent RC circuit, where the on-chip capacitance is represented by multiple narrow-band parallel RC circuits connected to the VDD and VSS power rails, see Fig.~\ref{fig:RLC_Model_freq_disp}. 
These circuits allow for an accurate representation of the PDN impedance over a wide frequency range.
The impedance characteristics of the PDN in the frequency domain provide valuable insight into the behavior of the system and can even be used to detect tampering events at the PCB level. 
Tampering with components inside the integrated circuit (IC), such as altering logic gates, placement, or routing, can change the on-chip capacitance, which in turn affects the PDN’s impedance.
This impact depends on the size, location, and nature of the tampering, demonstrating the sensitivity of the PDN to modifications within the chip~\cite{mosavirik2023silicon}.

\subsection{Interposer Power Distribution Network}
Today's PDN design for multi-chip modules is meticulously designed to manage the high power demands of large-scale compute circuits while maintaining signal integrity and effective thermal management. In this particular work, we focus on Xilinx/AMD's Stacked Silicon Interconnect (SSI) technology, which provides chiplet-based interconnections in multi-chip FPGAs.
At the core of this architecture is the silicon interposer, a passive layer that facilitates routing for configuration, global clocking, and interconnect signals between the programmable logic units known as Xilinx's Super Logic Regions (SLRs).
Each SLR in the SSI-enabled device operates with independent power delivery, with power and ground connections routed through the interposer.
These connections are facilitated by Through-Silicon Vias (TSVs), which create low-resistance paths between the active layers of the FPGA and the package substrate.
This segmentation minimizes cross-talk between regions, enhancing both power delivery efficiency and overall signal integrity.

The interposer enables the use of separate power planes for core logic, I/O, transceivers, and memory interfaces, reducing electrical noise and ensuring the stable operation of sensitive circuitry.
Furthermore, it distributes power uniformly across the SLRs and facilitates clocking and configuration connections while also contributing to thermal management.
At the chip level, the PDN forms a complex passive network that delivers power to each computing unit via the silicon interposer. 
The power supply transitions through interconnects between the package and the interposer layer before being locally distributed by the die power grid, an irregular, multi-layer metal mesh that connects to lower-level digital circuits.


\subsection{Coupling Effect Between SLRs}
The coupling effect in 2.5-D integrated systems occurs due to shared resources, such as the PDN and interposer, which interconnect multiple SLRs. 
This effect is driven by both power and electromagnetic interactions.
Variations in power consumption by one SLR induce fluctuations in the shared PDN, creating voltage and current perturbations that propagate across the interposer and impact neighboring SLRs. 
These fluctuations result from the resistive, capacitive, and inductive characteristics of the PDN, including mutual inductance and parasitic capacitance in the closely spaced power and ground planes, vias, and traces. 
Additionally, the switching activity in one SLR generates alternating electric and magnetic fields that can couple into adjacent SLRs through electromagnetic interference. 
These electromagnetic coupling signatures, while enabling SLRs interaction and optimization, also serve as reliable indicators for tamper detection and system security by revealing anomalies in power or electromagnetic behaviors~\cite{giechaskiel2019reading}.

\subsection{Delay-based On-Chip Sensor}

It has been shown in multiple cases that the voltage fluctuations due to computation on the IC's die could be measured with delay-based circuits placed on the same die~\cite{zhao2018fpga}.
Power consumption alteration affects the propagation delays of electrical signals.
Such change in propagation delay can be recorded in on-die sensory circuits, such as ring oscillators~\cite{gravellier2019high} and time-to-digital converters (TDCs)~\cite{zhao2018fpga}.
These sensors have been used to detect voltage, electromagnetic (EM), and laser glitching and radiating attacks on FPGAs~\cite{kajol2023ahd,monfared2024laserescape}.
However, they have not been used for integrity checks on multi-chip or chiplet systems.
Implementation of RO-based sensors incurs high power usage, and they perform poorly for accurate sensing (picosecond granularity).
Therefore, in this paper, we utilize TDC sensors for our integrity-checking framework. 
\begin{figure}[t]
\centering
\includegraphics[width=\columnwidth]{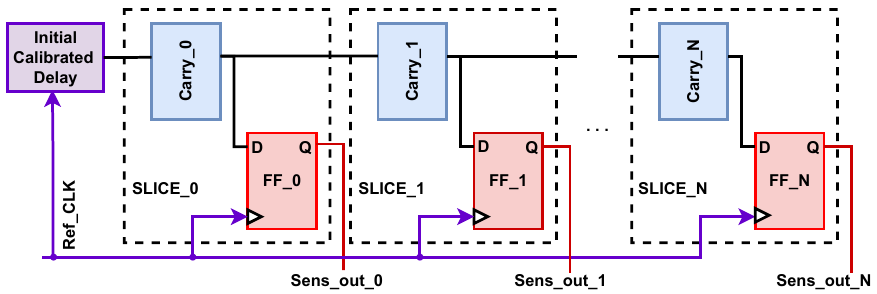}
\caption{high-level implementation of a TDC-based fault detection sensor}
\label{fig:tdc_sensor}
\end{figure}

In this work, inspired by the TDC implementation described by Gnad et al. ~\cite{gnad2021voltage}, we re-purpose a TDC design for integrity check purposes on multi-die FPGAs.
Fig.~\ref{fig:tdc_sensor} depicts a high-level implementation of the employed TDC.
As highlighted, the sensor's core components include an initial delayed signal, a tapped delay line, and corresponding output registers.
The calibration process is necessary for the TDC sensors and is performed offline by adjusting delay elements (e.g., CARRY4) or by modifying the chain of combinational logic blocks. 
This process ensures the sensor's output reaches a metastable state suitable for accurate sensing of tiny voltage fluctuations.
The delay line typically employs fast carry propagation logic, physically constrained to form a sequential delay chain.
Each delay unit's output is propagated to an output register clocked by the original signal, with the binary sensor output determined by the propagation delay at each register's input. 
Ideally, an external high-speed sampling circuitry is deployed to record the output of the TDC sensors. 
In this work, we organize a 2D mesh of TDC sensors on the verifier chiplet and perform a high-speed sampling on all TDCs simultaneously.


\section{ChipletQuake}
 \begin{figure*}[!t] 
\centering
\includegraphics[width=1\linewidth]{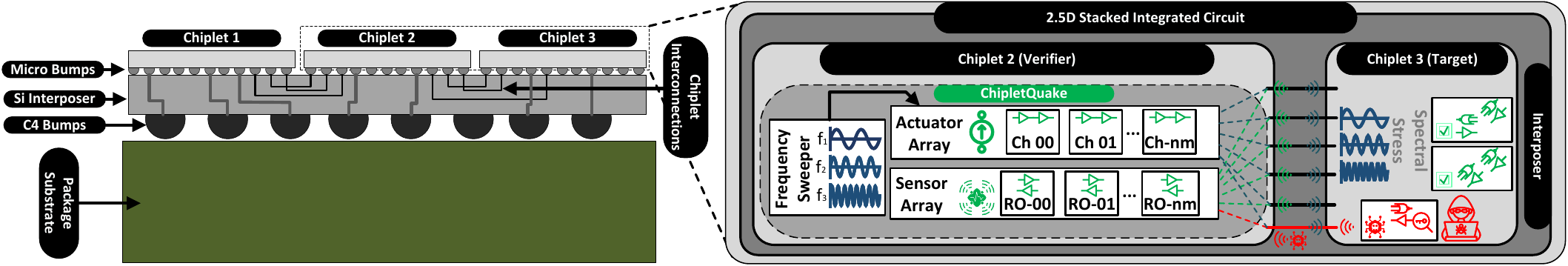}
\caption{High-level overview of \textsc{ChipletQuake} functionality}
\label{overview}
\end{figure*}

Considering the recent concerns regarding the potential chiplet-based security threats~\cite{juan2024hardware}, defense against such threats requires run-time and an accurate integrity-checking mechanism in place.
Motivated by the capacity of the shared PDNs in chiplet-enabled systems and the existing research for inter-chiplet fingerprinting~\cite{zhu2023pdnsig}, we develop a digital monitoring system to sense the impedance fluctuations of the physical environment of the chiplets.
We present \textsc{ChipletQuake} to detect tiny and passive alterations in the interposer, as well as neighboring chiplets, which can serve as a verification framework to identify Hardware Trojans (HT).

\subsection{High-level Design}
Fig.~\ref{overview} shows the high-level design of \textit{ChipletQuake}. 
As shown, the verifier chiplet is equipped with some internal components of a monitoring system.
With the use of a digitally designed frequency sweeper, an array of power waster circuity is activated by oscillating on certain frequencies. 
We refer to these elements (e.g., inverter chains) as \textit{Actuator Array}.
Due to the shared PDN in the system, neighboring chiplets undergo frequency-dependent current and voltage fluctuation.
At the same time, another sensory circuit in \textit{ChipletQuake} records this fluctuation by accurately sensing the voltage via digital delay-based or ring oscillator-based sensors. 

During the monitoring routine, if the verifier identifies any noticeable deviation from the valid existing measurement (golden model) on the sensed values, it considers a possible malicious hardware alteration on the neighboring chiplets or as shown on the interposer itself. 

\subsection{Chiplet Verification Flow }
\begin{figure}[!h]
\centering
\includegraphics[width=0.9\columnwidth]{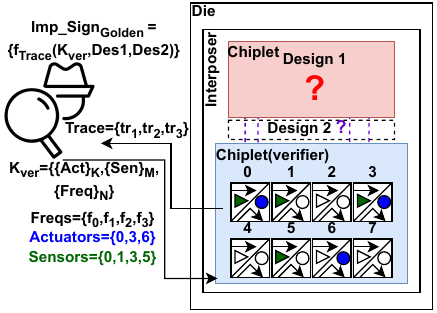}
\caption{Chiplet Verification Flow via Impedance Estimation }
\label{fig:flow}
\end{figure}

To facilitate a reliable hardware-level integrity check, we develop a verification protocol that utilizes a challenge-response authentication scheme~\cite{shoukry2015pycra}.
Instead of conventional digital signatures, a specific estimation of circuit impedance is considered as the response (signature).
As illustrated in Fig.~\ref{fig:flow}, the verifier chiplet deploys a 2D array on its reconfigurable fabric (e.g., FPGA).
Inside each element of the array located in the chiplet layout, a sensor (e.g., TDC) is implemented, which is highlighted as a filled \textcolor{deepgreen}{\raisebox{1.9ex}{\rotatebox{270}{\scalebox{1.6}{$\blacktriangle$}}}} if enabled during the verification process.

Furthermore, elements of the arrays are also equipped with a current actuator illustrated as a filled \textcolor{blue}{\raisebox{-0.2ex}{\scalebox{1.2}{\ding{108}}}} once activated.
Consequently, the verifier organizes a 2D grid of sensors and power wasters. 

To extract a proper impedance-based signature as the golden model, the verifier first chooses a set $N$ frequencies $\{Freq_{N}\}=\{f_0,f_1,f_2,... f_{N-1}\}$ in which the impedance is estimated during the verification process.  
Then, by choosing a set of corresponding IDs in the grid, $K$ actuator elements are selected.
For instance in Fig.~\ref{fig:flow}, actuators from the regions $\{Actuators\}=\{AC_0,AC_3,AC_6\}$ are selected and highlighted.
These actuators are then activated and fed with the selected frequencies from $\{Freq_{N}\}$ set. 
At the same time, the verifier selects another set of $M$ IDs, which are associated with the sensors that are responsible for recording the voltage-based fluctuations.
As a simple example in Fig.~\ref{fig:flow}, $\{Sensors\}=\{S_0,S_1,S_3,S_5\}$ are chosen. 

Performing the entire verification process locally on the verifier chiplet requires the verifier to store all the golden signatures on the verifier chip.
However, due to the possibility of storage limitation, the large size of multiple golden signatures makes this approach infeasible in some cases. 
Furthermore, to comply with the requirements of remote attestation scenarios~\cite{coker2011principles}, the integrity check should provide guarantees for remote users. 
In this case, a large attack surface should be considered. 
Specifically, reply attacks on the signatures can bypass the verification process.
Furthermore, dynamic reconfiguration of the target chiplets can affect the extracted signatures captured by the verifier. 
To overcome such challenges, \textit{ChipletQuake} provides a one-time authentication protocol by utilizing a one-time verification key $K_{ver}$ as the challenge for the target adjacent chiplets and the interposer itself.
The verification key is generated by randomly selecting sets of IDs of the \textit{Actuators}, \textit{Sensors}, and the target \textit{Frequencies} for the impedance sensing.
Our experiments aligned with the previous studies show that each combination of the $\{K_{ver}\}=\{\{Act\}_K,\{Sen\}_M,\{Freq\}_N\}$ yields a unique impedance profile which then can be post-processed to represent a valid golden signature. 

After recording multiple numbers of golden impedance signatures using different $\{K_{ver}\}$s, the target device in the field can be tested on demand. 
The traces extracted from the verification process can be easily validated remotely.
Note that, during the test process, the traces extracted from the chip's environment are a function dependent on $\{K_{ver}\}$ (which protects the security of the authentication protocol) and the physical layout and state of the hardware circuitry.
Hence, any malicious hardware-level modification on the system in a package can be detected by analyzing the captured traces.  

\subsection{Frequency Band Selection}
A PCB's power and ground planes form a cavity that can resonate at specific frequencies, occurring when the electrical dimensions of the PCB align with integer multiples of the electromagnetic wavelength. 
At these resonant frequencies, the impedance of the PCB becomes more sensitive due to the interaction of interconnect parasitic inductance, parasitic capacitance, and decoupling capacitance.
This intensified sensitivity makes the PCB more susceptible to even small changes in the impedance, such as those caused by a hardware Trojan and tampering events.
These small changes can introduce additional parasitics or alter signal pathways, potentially leading to observable performance anomalies.

Additionally, at resonant frequencies, electromagnetic energy tends to concentrate in certain regions of the PCB.
This localized accumulation of energy creates areas where the presence of a hardware Trojan can have a significant impact. 
A Trojan embedded in such regions may interact with the concentrated electromagnetic fields, leading to detectable deviations in signal integrity, power distribution, or electromagnetic emissions~\cite{han2018cipa}. 
The resonant frequency of a rectangular cavity resonator can be derived from the following equation:
\begin{equation}\label{resonance_cavity} f_r = \frac{1}{2 \pi \sqrt{\mu \epsilon}} \cdot \sqrt{\left( \frac{m \pi}{a} \right)^2 + \left( \frac{n \pi}{b} \right)^2 + \left( \frac{p \pi}{d} \right)^2} \end{equation}
where a and b are the cross-sectional sizes, and d is the length of the cavity while $\epsilon$ and $\mu$ are permittivity, and permeability of the material that cavity filled with it, respectively~\cite{ramo1994fields}.

Knowing the physical dimensions of the target PCB and its relative permittivity, the fundamental resonate frequencies can be estimated for measurements. 



\subsection{Implementation Layout}

\begin{figure}[t]
\centering
\includegraphics[width=0.9\columnwidth]{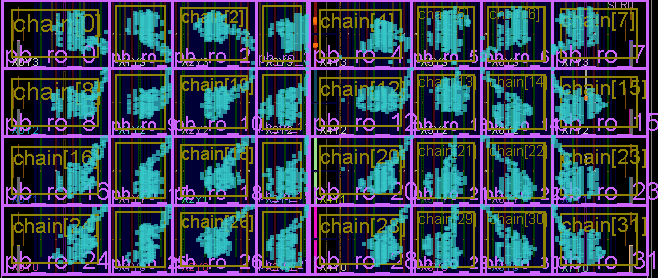}
\caption{\textsc{ChipletQuake} FPGA implementation layout with an array of 32 monitoring blocks}
\label{fig:layout}
\end{figure}

To implement \textit{ChipletQuake}, it is vital to properly carry out the physical placement of the underlying blocks. 
Specifically, we make physical constraints on the chiplet's fabric to ensure that sensors and actuators are evenly distributed and deployed. 
Fig.~\ref{fig:layout} shows an implementation layout of \textit{ChipletQuake} with 32 monitoring blocks on an FPGA chiplet.
As mentioned, each block comprises a TDC sensor and an inverter-based actuator.
Separated physically, each of these elements is controlled individually by a controller and can be configured to operate in different modes.
Specifically, the sampling rate of the sensors, input frequency of the actuator chain, and sensing time interval can be remotely configured as parameters during the verification procedure. 

\section{Case Studies}\label{sec:case}
\subsection{Threat Model}

The modern chiplet architectures necessitate the design teams to integrate or procure chiplet intellectual property (IP) from external vendors.
However, it is impractical for these design teams to be directly involved in the development of every individual chiplet.
Consequently, organizations that outsource chiplet components must depend on the manufacturers of these chiplets to deliver reliable hardware unless robust security measures are implemented.
In such a horizontal supply chain, to guarantee the quality and integrity of the products, outsourcing companies should adopt a Zero-Trust security model where no inherent trust is granted to devices or users, network environment, or ownership~\cite{stafford2020zero}. 
Applying a Zero-Trust approach demands the verification of all chiplet hardware, regardless of its source. 
Key vulnerabilities in chiplet-based systems include risks such as hardware tampering, unauthorized probing, and the insertion of hardware Trojans.
Given the extensive and global nature of the chiplet supply chain, authentication is critical to ensure that each chiplet meets the required standards and specifications.

To address these challenges, our fingerprinting method ensures the authenticity of the adjacent chiplets and the host interposer by exploiting digital impedance estimation. 
Furthermore, existing authentication mechanisms depend on keeping the challenge-response pair secure from adversaries, which requires that the key itself is stored within a trusted environment.
However, in the proposed method, the physical characteristics of the hardware account for the authenticity of the target hardware. 
Hence, any malicious tampering with the hardware yields an invalid key impeded in the impedance profile. 

In our threat model, we consider multi-chiplet system where a single challenger chiplet verifies the integrity of the adjacent chiplet and the interposer.
We assume no access to target chiplets in terms of logical challenge/response authentication.
Moreover, our threat model requires the verifier to store post-silicon golden signatures prior to the test procedure.
Furthermore, note that our treat model does not require the target chiplet circuitry to be activated in terms of transistor switching activity. 
This is a particularly useful model for detecting dormant and evasive hardware Trojans, which are often implemented via tiny units.

\subsection{Experimental Scenarios}

\begin{figure}[t!]
\centering
\includegraphics[width=0.95\columnwidth]{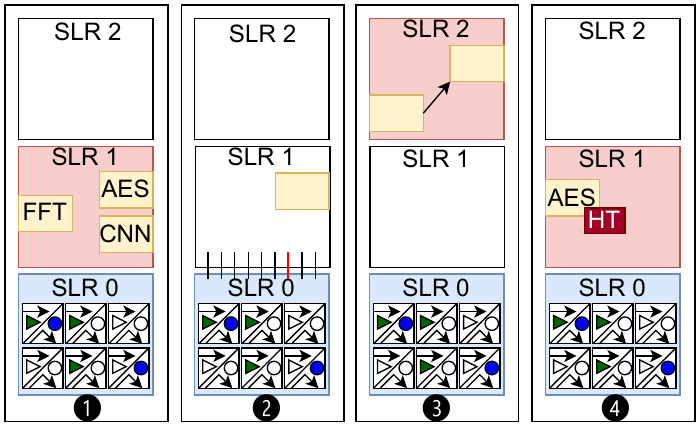}
\caption{High-level illustrations of the evaluation case studies }
\label{fig:case}
\end{figure}

We organize different case studies to thoroughly investigate the detection capability of the purposed digital impedance sensing in chiplet-based FPGAs.
Particularly, in this work, we focus on four distinct case studies for chiplet verification. Fig.~\ref{fig:case} highlights a high-level illustration of how each of these case studies is designed.
As shown, in each of the scenarios, we deploy our verification hardware in chiplet-0 (SLR0) and perform verification for chiplets/interposers.
We first study the case where different hardware modules are implemented in the adjacent SLR in \circled{1}. Here, we capture the impedance estimation for three hardware applications individually, and then we investigate if each of these hardware modules create a distinguishable fingerprint that can be detected by our sensors in the SLR0. 

In \circled{2}, we study if modifications on the host interposer can be effectively detected.
For this aim, we modify the utilization of communication lines between two chiplets.
These lines are implemented and placed on the host interposer in the FPGA. 
The scenario here emulates the attacks that include interposer tampering and probing. 
For the \circled{3} scenario, we further investigate the sensitivity of our framework for far chiplets. Specifically, we consider changing the placement of a single IP in SLR2 and monitoring the estimated profile on the verifier. 
Finally, in \circled{4}, we investigate if tiny hardware Trojans can be detected in adjacent chiplets.

\section{Evaluation Results}
\subsection{Implementation Setup}
For our evaluation, we used AMD Virtex™ UltraScale+™ VU37P HBM FPGA with 8GB of HBM DRAM memory. 
The target FPGA in this evaluation kit is \textit{XCVU37P-L2FSVH2892E}, which includes three programmable FPGA chiplets (i.e., SLRs) and corresponding HBM DRAM chiplet.
In our evaluations, we implement \textit{ChipletQuake} on \textit{SLR 0} as the verifier and perform the integrity/verification tests for other SLRs. 
Furthermore, the device is connected via a UART serial connection to the controller computer that receives the traces and generates the signature.

\subsection{Evaluation Metrics}
As a commonly used metric in the hardware security domain, we chose Welch’s 
\textit{t-test}~\cite{bilgin2014higher} as the basic distance metric for our evaluations. 
For Welch’s \textit{t-test}, small p values yield to reject the null hypothesis of similar (normal) distributions in distinguishability tests. 
For the sake of simplicity, it is best practice to usually select a threshold of $|t| > 4.5$ to reject the null hypothesis without considering the degree to conclude that the sets were drawn from different populations~\cite{schneider2015leakage}.
 
Nevertheless, as the impedance profile distribution might not necessarily follow a Gaussian trend in some cases, we also include the distribution-agnostic Wasserstein metric~\cite{arjovsky2017wasserstein} to carry out distinguishability tests.    
The Wasserstein metric is the function that provides the distance between two probability distributions, each extracted from impedance estimation via \textit{ChipletQuake}.
The $p^{\rm th}$ ($p\geq 1$) Wasserstein distance between $\gamma_{i}$ and $\tau_{i}$ is given by
\begin{equation}
    W_p(\gamma_{i},\tau_{i}) = [\textrm{inf}\ \mathbb{E}(d({{Im}^G_i},{{Im}^T_i}))^p]^{(1/p)}
\end{equation}

\noindent where $\mathbb{E}({Im})$ is the expected value of a random variable ${Im}$ (estimated Impedance in this case), $d$ is the Euclidean distance between two points, and the infimum is taken over all joint distributions of the random variables ${Im}^G_i$ and ${Im}^T_i$ with PDFs $\gamma_{i}$ and $\tau_{i}$, respectively.

\subsection{Profiling Different Hardware Designs}
\begin{figure}[h!]
\centering
\includegraphics[width=0.9\columnwidth]{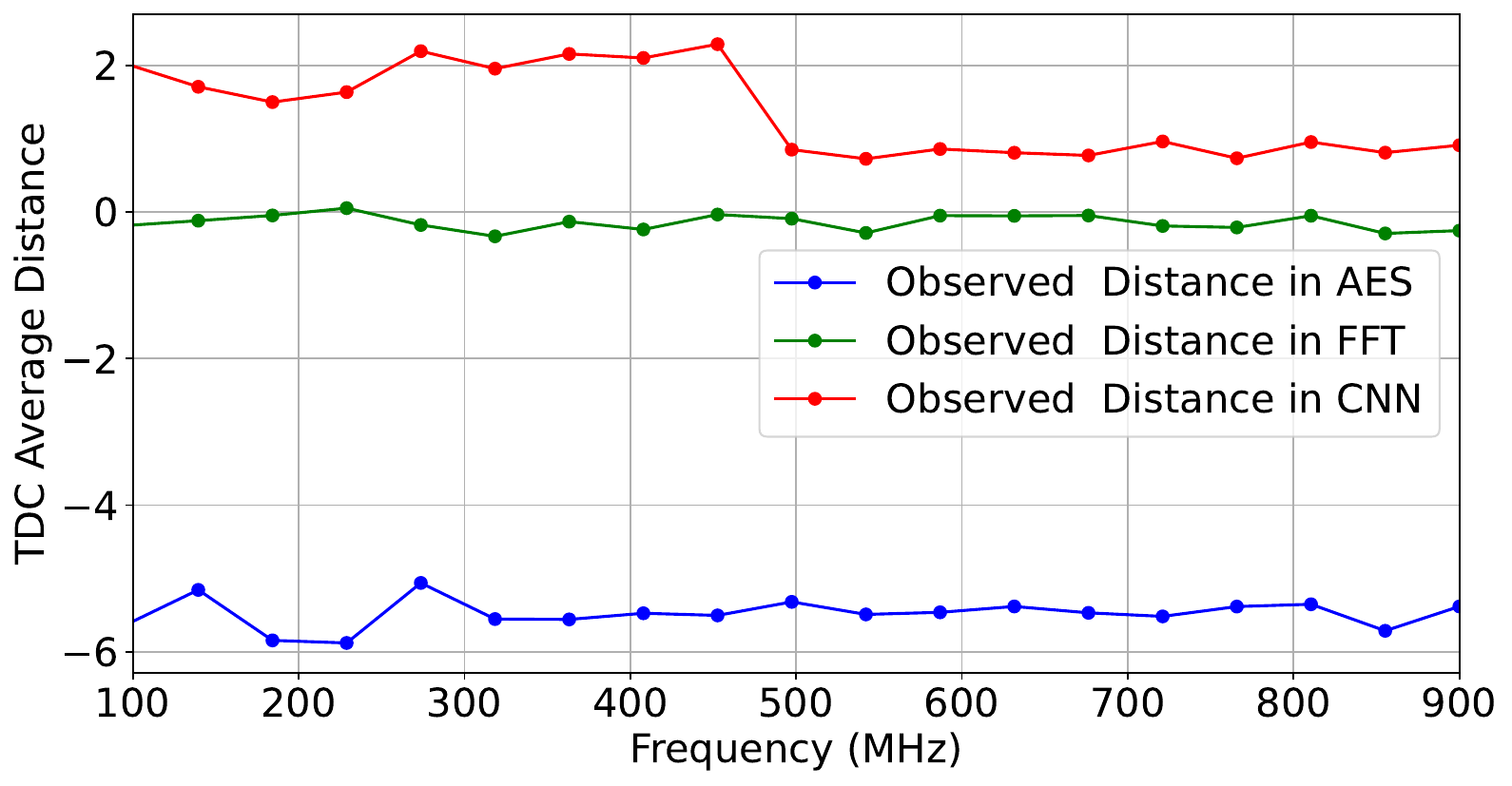}
\caption{TDC average distance on $T=500$ traces for different configurations}
\label{fig:diff_tdc}
\end{figure}

As for our first set of experiments (Case Study \circled{1}), we simply collect three sets of traces of $T=500$ for three hardware applications. Specifically, we consider AES, FFT, and a simple CNN, and then we calculate the average distance of the TDC sensed values with respect to a reference design on SLR1. Fig.~\ref{fig:diff_tdc} plots the average distance for each of these designs on the frequency point where the actuators were activated.


\subsection{Detecting Interposer Tampering}

\begin{figure}[t]
\centering
\includegraphics[width=0.9\columnwidth]{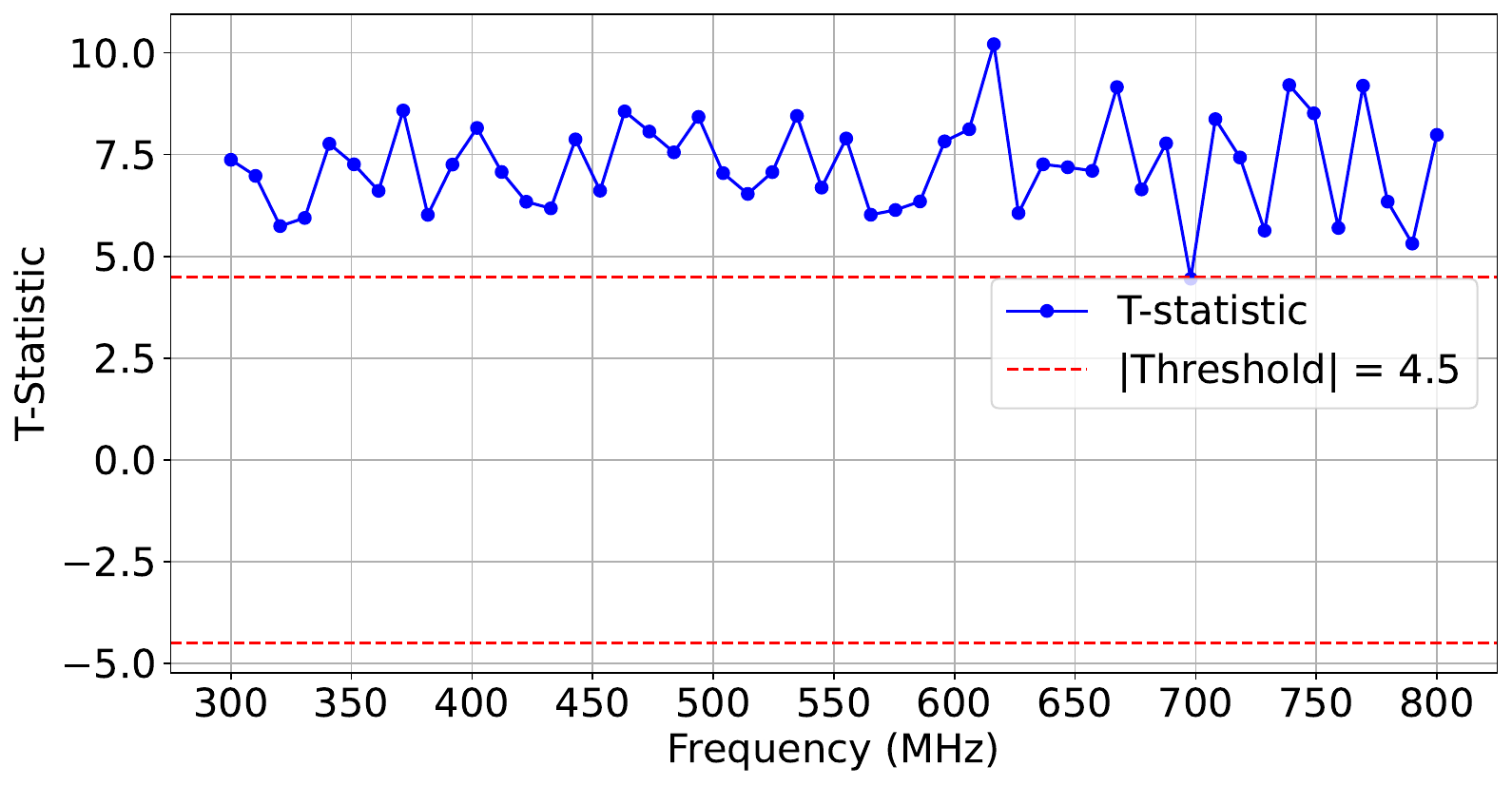}
\caption{T-test on $T=1000$ traces for SLL129 Vs SLL133}
\label{fig:ttest_ssl}
\end{figure}

\begin{figure}[t]
\centering
\includegraphics[width=0.9\columnwidth]{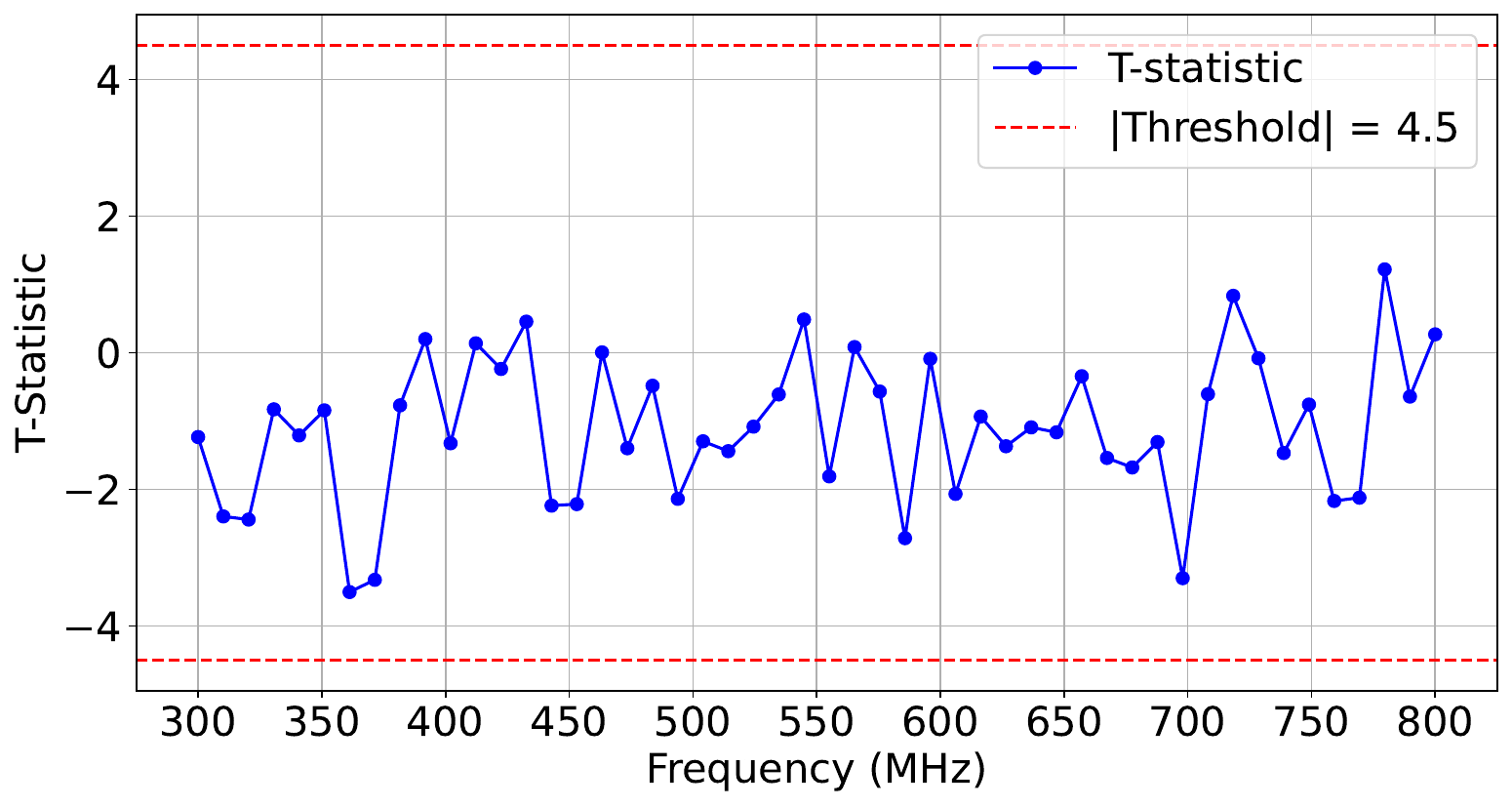}
\caption{T-test on $T=1000$ traces for SLL129 Vs SLL129-Ref}
\label{fig:ttest_ssl_ref}
\end{figure}

To verify the integrity of the interposer in our experiments, we generate two identical logical RTL hardware on \textit{SLR 1} as shown in \circled{2} in Fig.\ref{fig:case}.
To emulate the interposer tampering, we set a different number of interposer-built SLL interconnections in each design to change the interposer utilization layout. 
Specifically our designs employ \textit{129} and \textit{133} SLLs between \textit{SLR 0} and \textit{SLR 1}.
Similar to the procedure we took for HT detection, we collect three sets of traces ($T=1000$ for each set), including reference golden traces.
Fig.~\ref{fig:ttest_ssl} and Fig.~\ref{fig:ttest_ssl_ref} illustrate the t-test score for each of these two designs and the reliability reference score over the span of a selected frequency, respectively.  

Furthermore, we calculate a null hypothesis threshold in each frequency point based on the Wasserstein distance.
For Wasserstein distance, the null distributions are generated by performing $1000$ bootstrap sampling with respect to the reference set.
We set the significance value to $p-value=0.01$ to estimate the threshold for rejecting the null hypothesis.
Fig.~\ref{fig:wass_ssl} highlights the Wasserstein distance for interposer modification based on SLL utilization. 

\begin{figure}[!h]
\centering
\includegraphics[width=0.9\columnwidth]{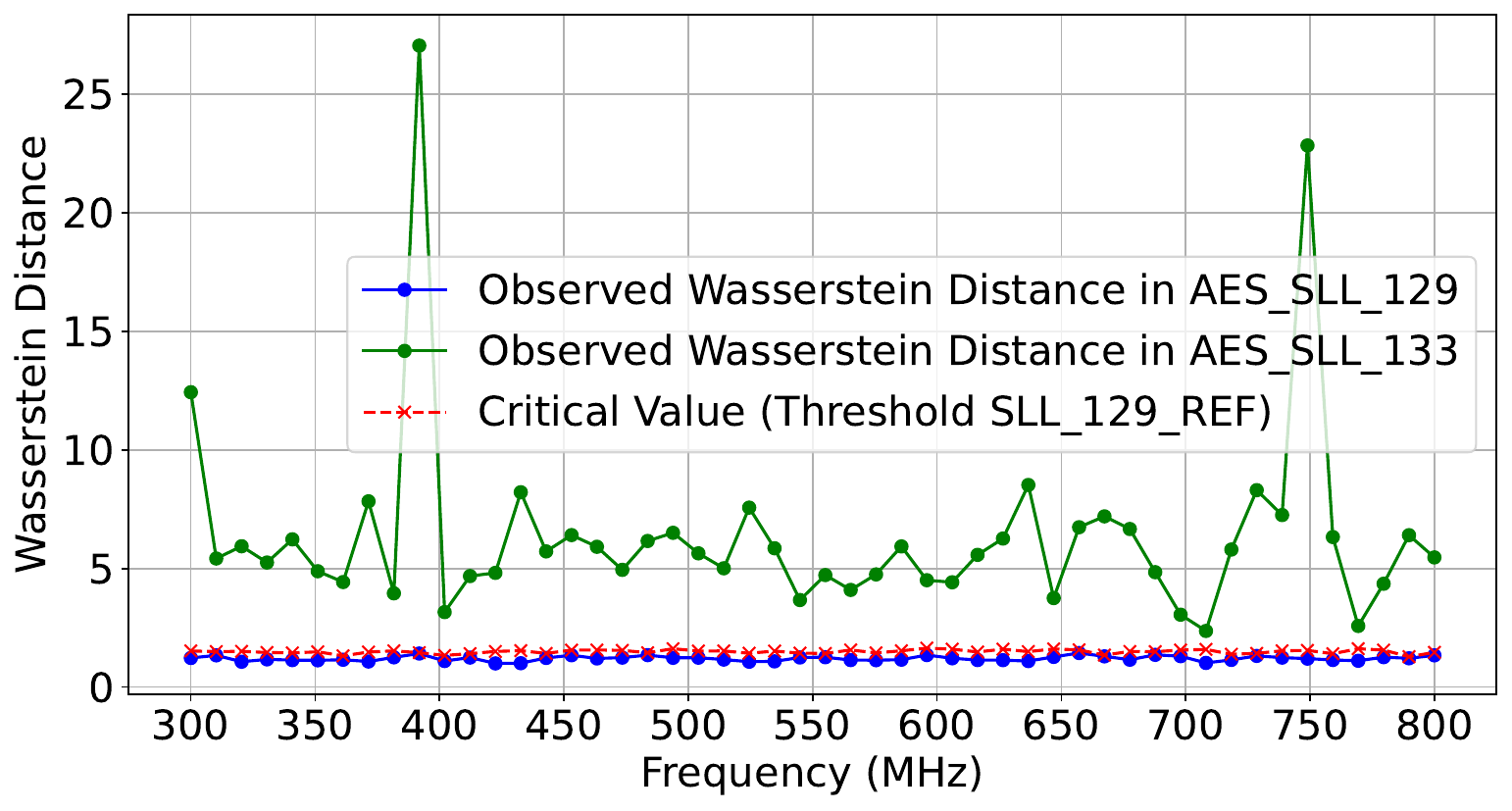}
\caption{Wasserstein distance on $T=1000$ traces for AES- SLL129 Vs SLL133}
\label{fig:wass_ssl}
\end{figure}

\subsection{Footprinting Further SLRs}
\begin{figure}[!h]
\centering
\includegraphics[width=0.9\columnwidth]{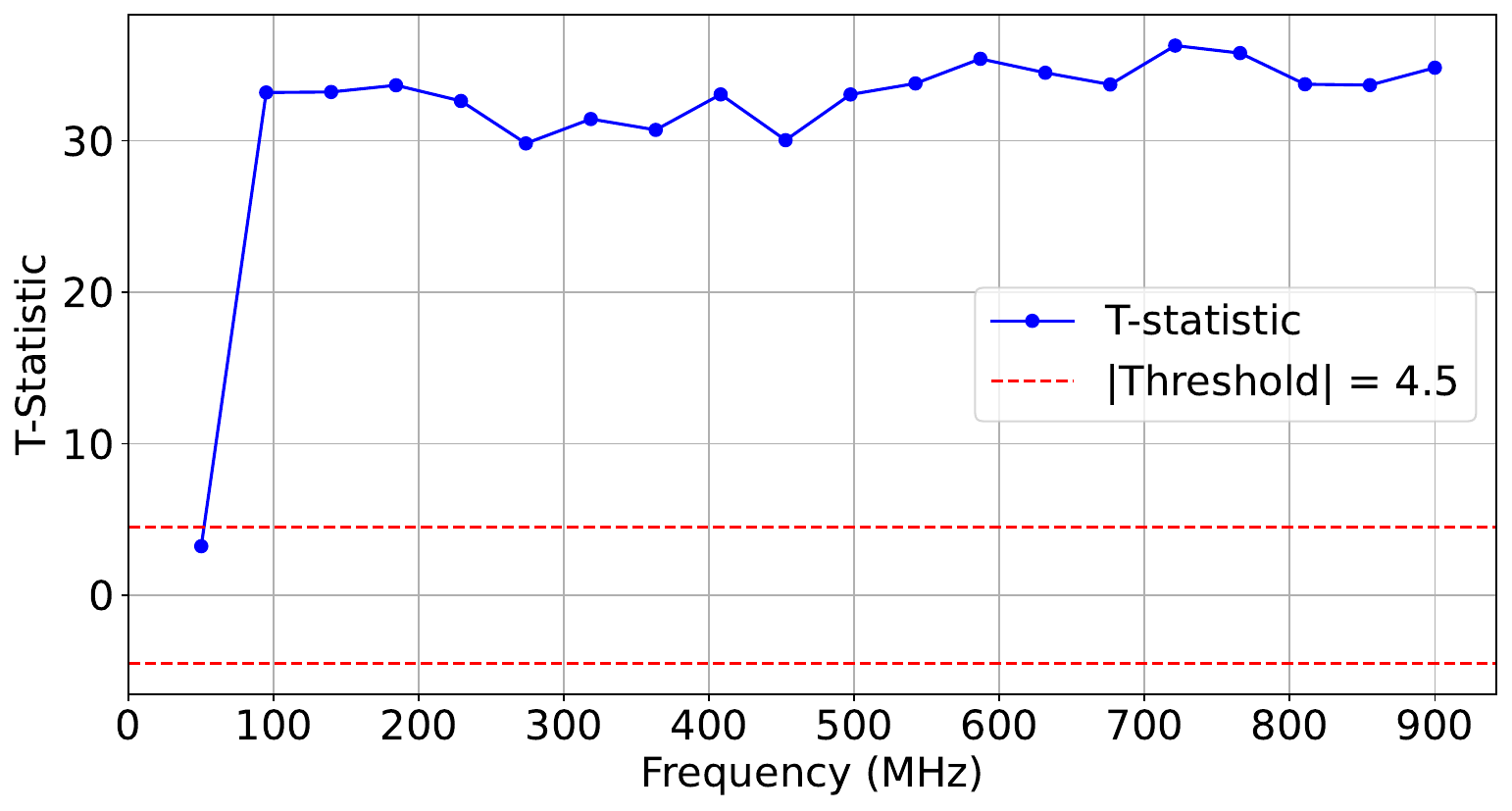}
\caption{T-test on $T=500$ traces for {$config_1$} Vs {$config_2$} at SLR2}
\label{fig:ttest_slr2}
\end{figure}

\begin{figure}[!h]
\centering
\includegraphics[width=0.9\columnwidth]{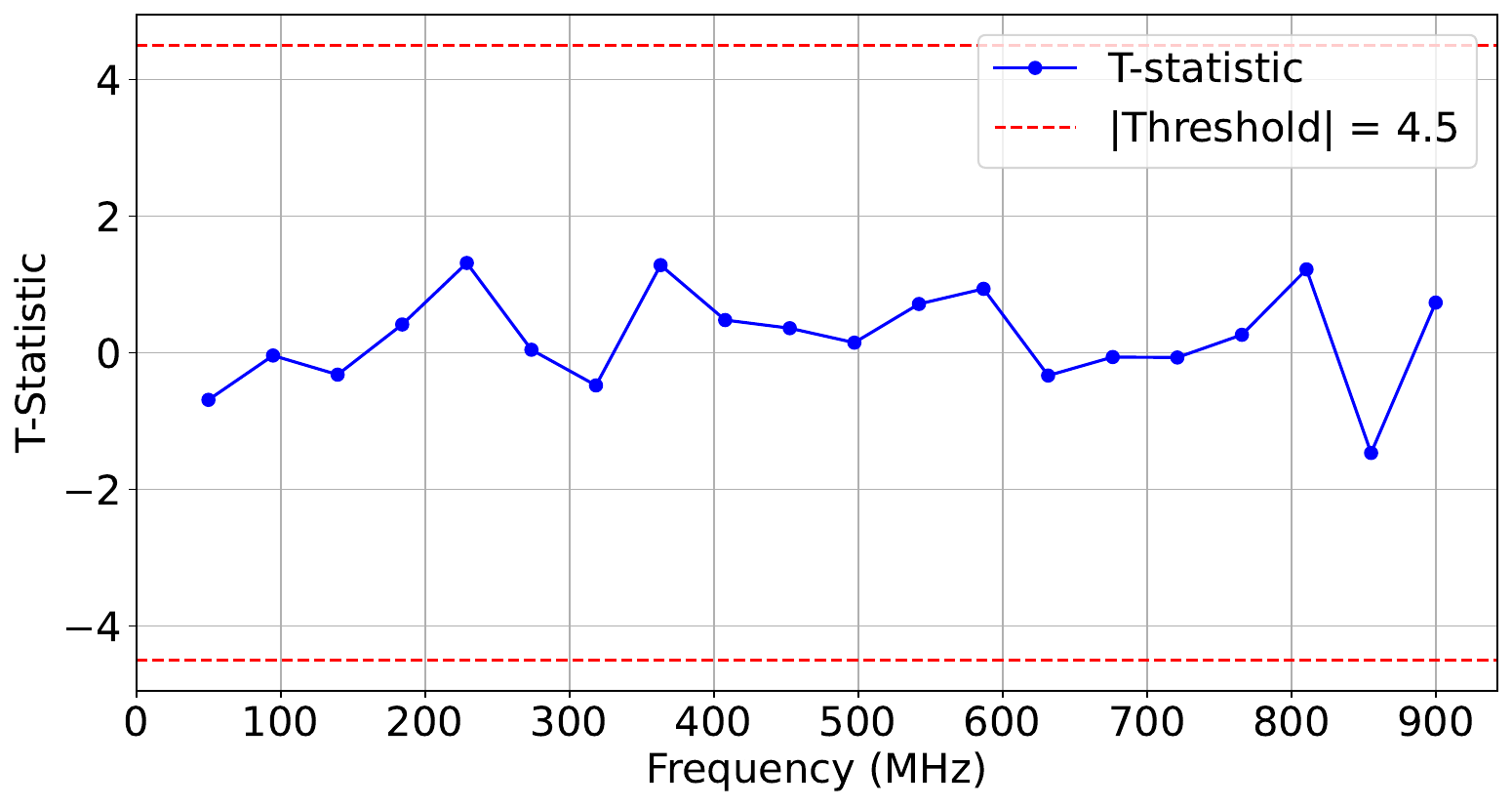}
\caption{T-test on $T=500$ traces for {$config_1$} Vs {$config_1$}-Ref On SLR2}
\label{fig:ttest_slr2_ref}
\end{figure}

Illustrated in \circled{3} in Fig.\ref{fig:case}, for this experiment, we implement a test hardware design on SLR2, which is fabricated at a farther distance from the verifier deployed on SLR0. 
We collect two sets of $T=500$ traces (and an additional reference set), where each set is associated with the same design in terms of functionality but routed differently in terms of implementation on the SLR2.
We refer to these designs as {$config_1$} and {$config_2$}.
Fig.~\ref{fig:ttest_slr2} and Fig.~\ref{fig:ttest_slr2_ref} showcase the t-test score for each of these two designs and the reliability reference score over the span of a selected frequency, respectively.

Moreover, Fig.~\ref{fig:wass_ssl} highlights the Wasserstein distance for between {$config_1$} and {$config_2$} with the {$config_1$} as the reference.  

\begin{figure}[!h]
\centering
\includegraphics[width=0.9\columnwidth]{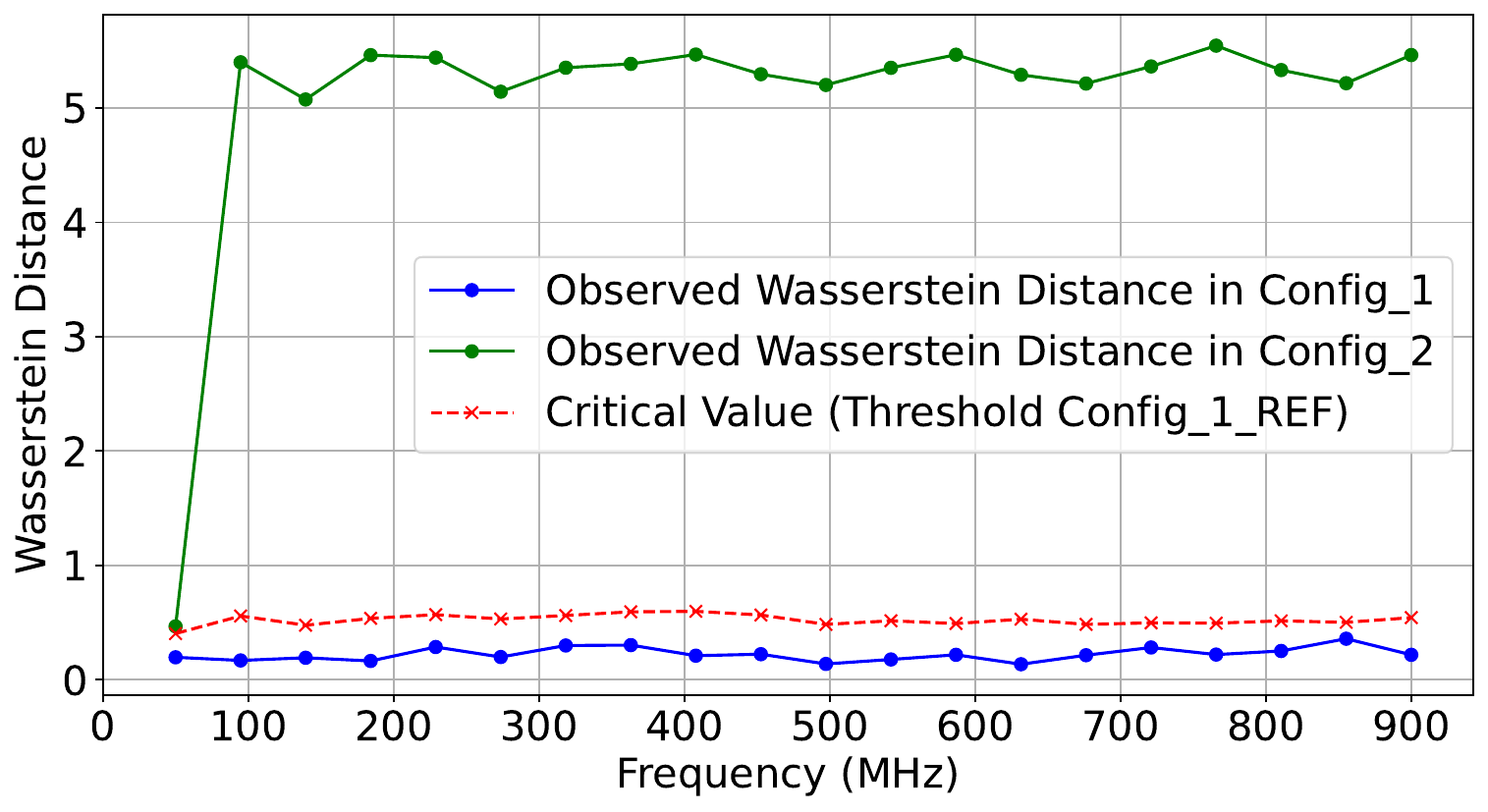}
\caption{Wasserstein distance on $T=500$ traces for {$config_1$} Vs {$config_2$} On SLR2}
\label{fig:wass_ssl}
\end{figure}

\subsection{Detecting Hardware Trojan}
In our final set of experiments, as highlighted in \circled{4}, we evaluate \textit{ChipletQuake} against the implementation of Hardware Trojans.
For this case study, we utilized the
Register-Transfer Level (RTL) HT benchmarks of AES implementations from Trust-Hub~\cite{thrusthub}. 
The original HT-free design in these implementations is an AES-128 block cipher IP, with an 11-stage pipeline to perform the ten stages of AES encryption on the 128-bit block.
For the HT implementation, we used AES-T1100 variation with an HT payload that modulates AES activity to create a power consumption pattern that leaks the AES key. 
We deployed each of these implementations on \textit{SLR 1} and performed a distinguishability test from the verifier on \textit{SLR 0} empowered by \textit{ChipletQuake}.

For our experiment, we collect three sets of traces of $T=500$, a set of traces for the legit $AES\_HT\_FREE$ hardware as the reference, another set of traces for $AES\_HT\_FREE$ for testing, and a set of malicious $AES\_HT$ implementation. 

\begin{figure}[h!]
\centering
\includegraphics[width=0.9\columnwidth]{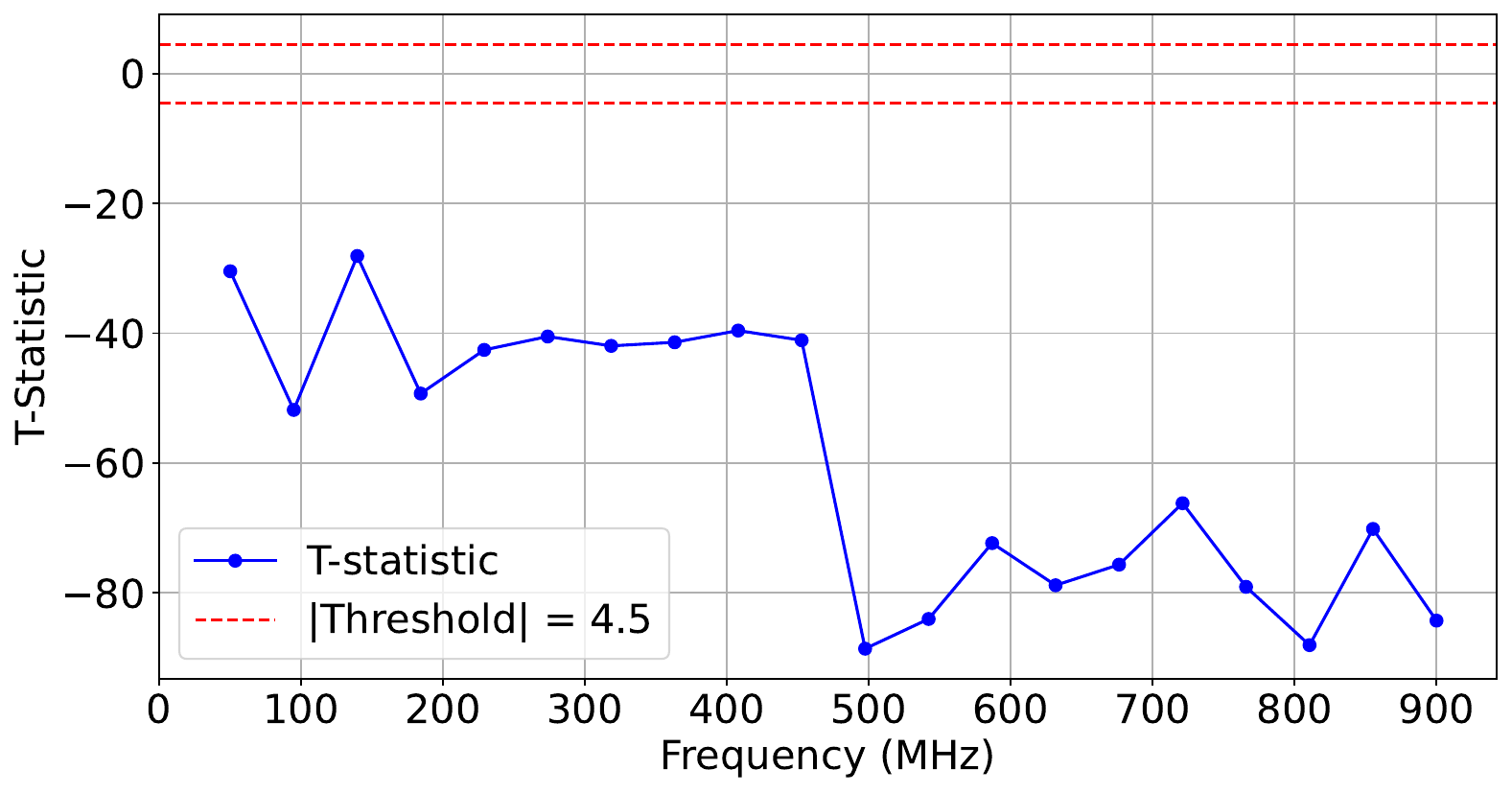}
\caption{T-test on $T=500$ traces for AES-HT-Free Vs AES-HT}
\label{fig:htvsfree}
\end{figure}

Fig.~\ref{fig:htvsfree} shows the frequency-based T-test distinguishably results captured for two sets of experiments where $AES\_HT\_FREE$ and $AES\_HT$ are implemented in the adjacent \textit{SLR 1}.

\begin{figure}[!h]
\centering
\includegraphics[width=0.9\columnwidth]{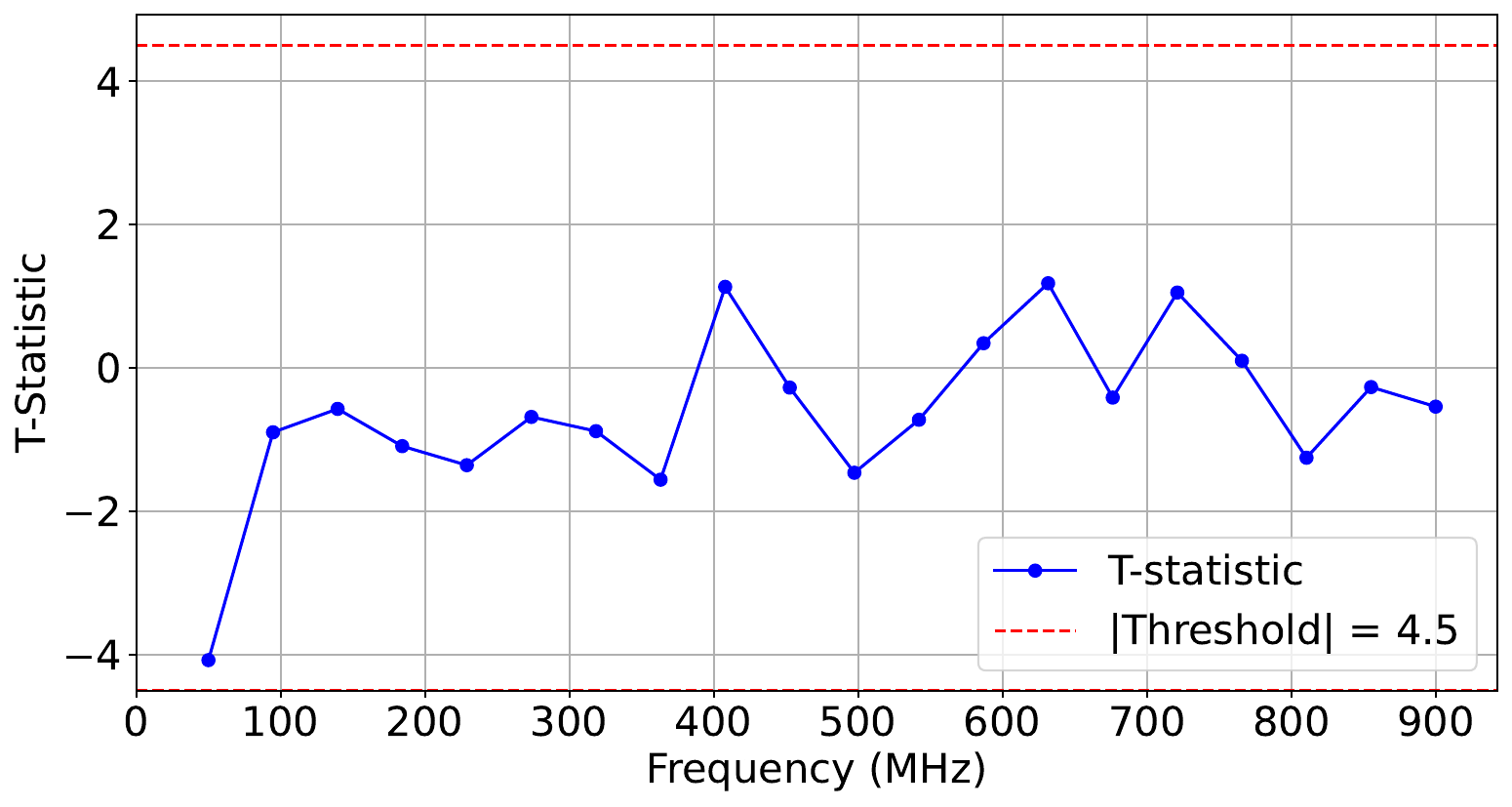}
\caption{T-test on $T=500$ traces for AES-HT-Free Vs AES-HT-Free-Ref}
\label{fig:free_ref}
\end{figure}

Furthermore, to showcase the reliability, we perform another set of HT-Free impedance estimation and compare it to the existing HT-Free reference traces.
Fig.~\ref{fig:free_ref}, depicts the T-test for two sets of identical implementation of HT-free AES.
As anticipated, we can verify the difference score is confined within the $|t|< 4.5$ similarity range.   

\begin{figure}[h!]
\centering
\includegraphics[width=0.9\columnwidth]{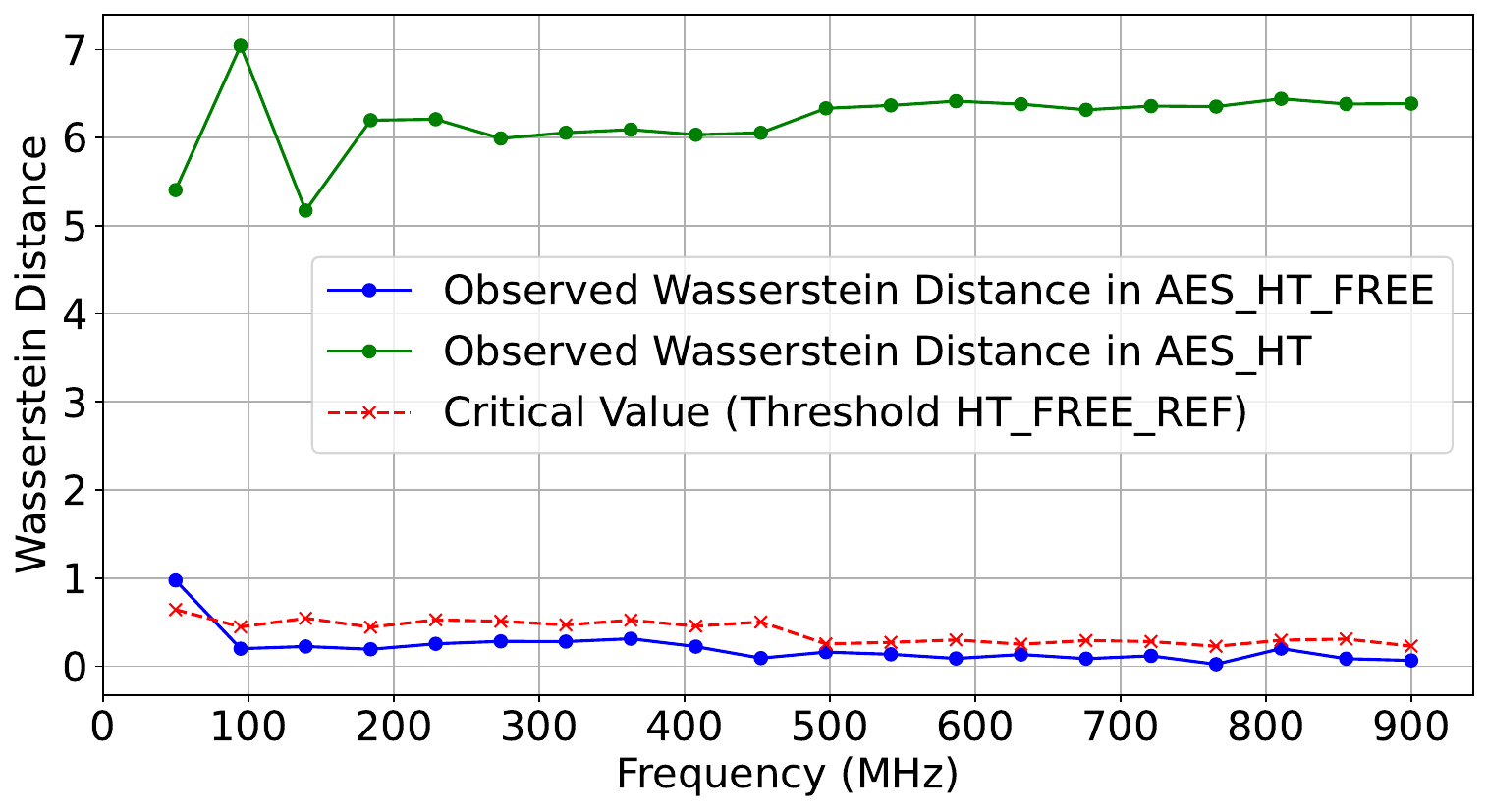}
\caption{Wasserstein distance on $T=500$ traces for AES-HT-Free Vs AES-HT}
\label{fig:wass_ht}
\end{figure}

As shown in Fig.~\ref{fig:wass_ht}, the calculated Wasserstein distance of the HT-included IP is visible by a large margin, and hence malicious circuit on the chiplet can be detected effectively.

\section{Discussions}\label{sec:dis}


\subsection{Comparison to Related Works}
Power-based side-channel attacks have been shown to be effective in chiplet-based systems to extract information from adjacent chiplets. 
In this regard, similar approaches such as~\cite{zhang2023sipguard} can be used to detect anomalies in adjacent chiplets.
Such works exploit the alteration sensed in power consumption of the neighboring chiplets due to the dynamic execution of the target circuits. 
Hence, static and dormant malicious circuits (e.g., HTs) are not effectively discovered unless activated.

There have also been similar works to utilize sensors in chiplets for verification and protection in SiP. Specifically, authors in~\cite{zhu2023pdnsig} use the impedance estimation via FFT to fingerprint the chiplets.
Although similar actuator/sensor architecture can be used to estimate impedance via voltage fluctuations, time-to-frequency conversion via FFT potentially drops the SNR required for intra-SLR fingerprinting.

Our work presents a frequency domain sweeping approach that provides a rich profile of impedance that can be processed to identify static circuits which are not required to be dynamically activated. 
Such methodology presents enough accuracy to detect tiny static dormant HTs.

\subsection{Further Applications}
It might be possible to use this framework to perform dynamic verifications as well. 
This means that the signature procedure can be performed during the switching activities of the adjacent chiplet.
The impedance-based dynamic fingerprinting potentially opens up the possibility of carrying out adversary side-channel attacks on the neighboring chiplets.
Furthermore, as another application, it is also possible to perform extensive templating/profiling to fingerprint a target IP in threat models that physical side-channels are considered. 



\section{Conclusion}
The transition to chiplet-based designs addresses critical challenges in modern semiconductor manufacturing but also introduces significant security vulnerabilities in the hardware supply chain.
This paper presented \textit{ChipletQuake}, a framework for post-silicon verification of physical security in chiplet-based systems.
By leveraging impedance sensing of the power delivery network (PDN), \textit{ChipletQuake} effectively detects tampering events in the interposer and neighboring chiplets without requiring additional hardware. The impedance estimation method is able to detect dormant, static, and tiny modifications on the MCMs.
Our experimental results demonstrate its capability to identify hardware Trojans and interposer tampering, validating its effectiveness in enhancing the security of FPGA-based chiplet systems. 






\bibliographystyle{IEEEtran}
\bibliography{ref}

\begin{thebibliography}{10}
\providecommand{\url}[1]{#1}
\csname url@samestyle\endcsname
\providecommand{\newblock}{\relax}
\providecommand{\bibinfo}[2]{#2}
\providecommand{\BIBentrySTDinterwordspacing}{\spaceskip=0pt\relax}
\providecommand{\BIBentryALTinterwordstretchfactor}{4}
\providecommand{\BIBentryALTinterwordspacing}{\spaceskip=\fontdimen2\font plus
\BIBentryALTinterwordstretchfactor\fontdimen3\font minus \fontdimen4\font\relax}
\providecommand{\BIBforeignlanguage}[2]{{%
\expandafter\ifx\csname l@#1\endcsname\relax
\typeout{** WARNING: IEEEtran.bst: No hyphenation pattern has been}%
\typeout{** loaded for the language `#1'. Using the pattern for}%
\typeout{** the default language instead.}%
\else
\language=\csname l@#1\endcsname
\fi
#2}}
\providecommand{\BIBdecl}{\relax}
\BIBdecl

\bibitem{deric2022know}
A.~Deric and D.~Holcomb, ``Know time to die--integrity checking for zero trust chiplet-based systems using between-die delay pufs,'' \emph{IACR Transactions on Cryptographic Hardware and Embedded Systems}, pp. 391--412, 2022.

\bibitem{deric2024evaluating}
A.~Deric, K.~Mitard, S.~Tajik, and D.~Holcomb, ``Evaluating vulnerability of chiplet-based systems to contactless probing techniques,'' \emph{arXiv preprint arXiv:2405.14821}, 2024.

\bibitem{zhang2023sipguard}
T.~Zhang, M.~L. Rahman, H.~M. Kamali, K.~Z. Azar, and F.~Farahmandi, ``Sipguard: run-time system-in-package security monitoring via power noise variation,'' \emph{IEEE Transactions on Very Large Scale Integration (VLSI) Systems}, 2023.

\bibitem{hossen2022analysis}
M.~O. Hossen, A.~Kaul, E.~Nurvitadhi, M.~D. Pant, R.~Gutala, A.~Dasu, and M.~S. Bakir, ``Analysis of power delivery network (pdn) in bridge-chips for 2.5-d heterogeneous integration,'' \emph{IEEE Transactions on Components, Packaging and Manufacturing Technology}, vol.~12, no.~11, pp. 1824--1831, 2022.

\bibitem{mosavirik2023silicon}
T.~Mosavirik, S.~K. Monfared, M.~S. Safa, and S.~Tajik, ``Silicon echoes: Non-invasive trojan and tamper detection using frequency-selective impedance analysis,'' \emph{IACR Transactions on Cryptographic Hardware and Embedded Systems}, vol. 2023, no.~4, pp. 238--261, 2023.

\bibitem{zhu2023pdnsig}
H.~Zhu, W.~Cao, and X.~Zhang, ``Pdnsig: Identifying multi-tenant cloud fpgas with power distribution network-based signatures,'' in \emph{2023 IEEE/ACM International Conference on Computer Aided Design (ICCAD)}.\hskip 1em plus 0.5em minus 0.4em\relax IEEE, 2023, pp. 1--8.

\bibitem{giechaskiel2019reading}
I.~Giechaskiel, K.~Rasmussen, and J.~Szefer, ``Reading between the dies: Cross-slr covert channels on multi-tenant cloud fpgas,'' in \emph{2019 IEEE 37th International Conference on Computer Design (ICCD)}.\hskip 1em plus 0.5em minus 0.4em\relax IEEE, 2019, pp. 1--10.

\bibitem{zhao2018fpga}
M.~Zhao and G.~E. Suh, ``Fpga-based remote power side-channel attacks,'' in \emph{2018 IEEE Symposium on Security and Privacy (SP)}.\hskip 1em plus 0.5em minus 0.4em\relax IEEE, 2018, pp. 229--244.

\bibitem{gravellier2019high}
J.~Gravellier, J.-M. Dutertre, Y.~Teglia, and P.~Loubet-Moundi, ``High-speed ring oscillator based sensors for remote side-channel attacks on fpgas,'' in \emph{2019 International conference on ReConFigurable computing and FPGAs (ReConFig)}.\hskip 1em plus 0.5em minus 0.4em\relax IEEE, 2019, pp. 1--8.

\bibitem{kajol2023ahd}
M.~A. Kajol, S.~Sunkavilli, and Q.~Yu, ``Ahd-lam: A new mitigation method against voltage-drop attacks in multi-tenant fpgas,'' in \emph{2023 Asian Hardware Oriented Security and Trust Symposium (AsianHOST)}.\hskip 1em plus 0.5em minus 0.4em\relax IEEE, 2023, pp. 1--6.

\bibitem{monfared2024laserescape}
S.~K. Monfared, K.~Mitard, A.~Cannon, D.~Forte, and S.~Tajik, ``{LaserEscape: Detecting and Mitigating Optical Probing Attacks},'' in \emph{2023 IEEE/ACM International Conference on Computer Aided Design (ICCAD)}, 2024.

\bibitem{gnad2021voltage}
D.~R. Gnad, C.~D.~K. Nguyen, S.~H. Gillani, and M.~B. Tahoori, ``Voltage-based covert channels using fpgas,'' \emph{ACM Transactions on Design Automation of Electronic Systems (TODAES)}, vol.~26, no.~6, pp. 1--25, 2021.

\bibitem{juan2024hardware}
S.~Juan, A.~Fady, P.~Anthony, R.~Philippe \emph{et~al.}, ``On hardware security and trust for chiplet-based 2.5 d and 3d ics: Challenges and innovations,'' \emph{IEEE Access}, 2024.

\bibitem{shoukry2015pycra}
Y.~Shoukry, P.~Martin, Y.~Yona, S.~Diggavi, and M.~Srivastava, ``Pycra: Physical challenge-response authentication for active sensors under spoofing attacks,'' in \emph{Proceedings of the 22nd ACM SIGSAC Conference on Computer and Communications Security}, 2015, pp. 1004--1015.

\bibitem{coker2011principles}
G.~Coker, J.~Guttman, P.~Loscocco, A.~Herzog, J.~Millen, B.~O’Hanlon, J.~Ramsdell, A.~Segall, J.~Sheehy, and B.~Sniffen, ``Principles of remote attestation,'' \emph{International Journal of Information Security}, vol.~10, pp. 63--81, 2011.

\bibitem{han2018cipa}
Y.~Han, X.~Wang, and M.~Tehranipoor, ``Cipa: Concurrent ic and pcb authentication using on-chip ring oscillator array,'' in \emph{2018 IEEE 27th Asian Test Symposium (ATS)}.\hskip 1em plus 0.5em minus 0.4em\relax IEEE, 2018, pp. 109--114.

\bibitem{ramo1994fields}
S.~Ramo, J.~R. Whinnery, and T.~Van~Duzer, \emph{Fields and waves in communication electronics}.\hskip 1em plus 0.5em minus 0.4em\relax John Wiley \& Sons, 1994.

\bibitem{stafford2020zero}
V.~Stafford, ``Zero trust architecture,'' \emph{NIST special publication}, vol. 800, p. 207, 2020.

\bibitem{bilgin2014higher}
B.~Bilgin, B.~Gierlichs, S.~Nikova, V.~Nikov, and V.~Rijmen, ``Higher-order threshold implementations,'' in \emph{Advances in Cryptology--ASIACRYPT 2014: 20th International Conference on the Theory and Application of Cryptology and Information Security, Kaoshiung, Taiwan, ROC, December 7-11, 2014, Proceedings, Part II 20}.\hskip 1em plus 0.5em minus 0.4em\relax Springer, 2014, pp. 326--343.

\bibitem{schneider2015leakage}
T.~Schneider and A.~Moradi, ``Leakage assessment methodology: A clear roadmap for side-channel evaluations,'' in \emph{Cryptographic Hardware and Embedded Systems--CHES 2015: 17th International Workshop, Saint-Malo, France, September 13-16, 2015, Proceedings 17}.\hskip 1em plus 0.5em minus 0.4em\relax Springer, 2015, pp. 495--513.

\bibitem{arjovsky2017wasserstein}
M.~Arjovsky, S.~Chintala, and L.~Bottou, ``Wasserstein generative adversarial networks,'' in \emph{International conference on machine learning}.\hskip 1em plus 0.5em minus 0.4em\relax PMLR, 2017, pp. 214--223.

\bibitem{thrusthub}
Trust-Hub, ``{Hardware Trojan Benchmarks},'' \url{[Online] https://trust-hub.org }, 2025.

\end{thebibliography}


\end{document}